\documentstyle[preprint,12pt,aps,epsfig]{revtex}  
\tightenlines
\begin{document}
\draft

\bigskip

\title{Polarized $J/\psi$ Production at CLEO}
\author{Seungwon Baek\thanks{Electronic address: 
swbaek@phya.snu.ac.kr}$^{(a)}$,
P. Ko\thanks{Electronic address: pko@chep6.kaist.ac.kr}$^{(b)}$ ,
Jungil Lee\thanks{
Address after Dec. 1, 1997:
Dep. of Physics, Ohio State Univ., Columbus, OH 43210, USA.
}$^{(a)}$, and
H.S. Song\thanks{Electronic address: hssong@physs.snu.ac.kr}$^{(a)} $}
\address{$^{(a)}$
Center for Theoretical Physics and
Department of Physics,\\ Seoul National University,
Seoul 151-742, Korea \\
$^{(b)}$
Department of Physics, KAIST, Taejon 305-701, Korea
}
\maketitle
\tighten
\begin{abstract}
Polarizations of the prompt $J/\psi$'s produced in the  $e^+e^-$ annihilation
at CLEO energy ($\sqrt{s} = 10.58$ GeV) are considered in the framework of 
NRQCD factorization  formalism. We find that the $J/\psi$ polarization
has strong dependence on the production mechanism. At CLEO energy, 
the most dominant $J/\psi$ production mechanism in the most phase space is 
the  color-singlet mechanism, $e^+e^- \rightarrow J/\psi+gg$, for which 
$J/\psi$'s  are highly longitudinally polarized.
On the other hand, the dominant $J/\psi$ production mechanism at the
upper end point  of $J/\psi$ energy distribution is the color-octet 
process, $e^+e^- \rightarrow (c\bar{c})^{(8)} +g$, for which  $J/\psi$'s
are almost unpolarized.  Thus, the measurement of the polarization of the 
end point $J/\psi$'s can give another test of color-octet mechanism, and 
constraint on the NRQCD matrix elements.
\end{abstract} 

\newpage

The nonrelativistic QCD(NRQCD)\cite{BBL} is an effective field theory 
of QCD that describes heavy quarkonium physics. It has a distinctive feature
that allows color-octet mechanism in heavy quarkonium production and decay.
After Braaten and Fleming suggested the color-octet mechanism as a possible 
solution to the $\psi^\prime$ anomaly at the Tevatron \cite{BF}, this idea 
has given possible explanations on several experimental data which could not 
be explained on the basis of color-singlet model (CSM) \cite{ko} \cite{z0}.
One crucial feature of the NRQCD approach is that a set of universal 
nonperturbative NRQCD matrix elements describes vastly different heavy
quarkonium production and decay processes. These nonperturbative parameters 
can be extracted from some experimental data or from lattice simulations 
\cite{lattice}, and then they can be used at other procedures.  Thus, one 
can check consistency of the whole approach based on the NRQCD factorization 
formalism. In view of this, it is important to  calculate as many
independent heavy quarkonium production and decay  processes as 
possible,  and see if  one can have a consistent picture of overall  
phenomenology  for heavy quarkonium physics.  
 
One such process is the inclusive $J/\psi$ production in the $e^+ e^-$ 
annihilation, $e^+e^-\to J/\psi + X$, which has been studied in 
various frameworks. Fritzsch and K\"uhn \cite{Fr} studied the process 
$e^+e^-\to J/\psi + g$ as the leading order contribution to the inclusive 
$J/\psi$ production subprocess in the color evaporation model.  This process
was considered in the CSM by various authors \cite{earlier}. Braaten and Chen 
\cite{Bra} showed that the color-octet contribution may dominate near the 
upper end point of the $J/\psi$ energy spectrum. In this region, the $J/\psi$ 
angular distribution can change dramatically due to the color-octet 
mechanism. Cho and Leibovich studied this process via color-singlet mechanism
in the NRQCD including complete $\alpha_s^2$ correction \cite{cho}. 
And Yuan, Qiao and Chao studied this process via both color-singlet and 
color-octet  mechanisms, and tried to  extract the color-octet matrix 
elements\cite{fyuan}.  Finally, other related processes such as $e^+ e^- 
\rightarrow J/\psi + \gamma$ and $e^+ e^- \rightarrow J/\psi + e^+ e^-$ 
were  considered by Chang {\it et al.} \cite{chang}.

In this paper, we consider the $J/\psi$ polarization in the $e^+e^-$
annihilation at the CLEO energy ($\sqrt{s} = 10.58$ GeV). 
The longitudinal polarization of the prompt $J/\psi$ is expected to be 
$42\%$ when the color-octet $c\bar{c}[{^1S_0} ~{\rm or}~ {^3P_J}]^{(8)}$ 
contributions are included, whereas the CSM alone predicts $53\%$. 
We show that  the energy dependence  of the longitudinal polarization is 
sensitive to  the $J/\psi$ production  mechanism. If the color-octet 
mechanism dominates at the upper end point of phase space, almost 
unpolarized $J/\psi$'s are produced.  If the color-singlet mechanism 
dominates, almost longitudinally polarized  $J/\psi$'s are produced.

In order to study  the polarization of $J/\psi$'s produced in the $e^+ e^-$ 
annihilation,  we define an observable $\eta(z)$ as
\begin{eqnarray}
\eta(z)\equiv\frac{d\sigma_{L}}{dz}\bigg/\frac{d\sigma}{dz}
\;,
\end{eqnarray}   
where $z \equiv  2 E_{J/\psi} / \sqrt{s}$,  $d\sigma_{L}/{dz}$ represents 
the energy spectrum of the logitudianlly polarized prompt $J/\psi$, 
and $d\sigma/{dz}$ that of total $J/\psi$ production.  Polarization of 
produced $J/\psi$'s can be obtained by measuring the angular
distribution of lepton pair in the subsequent decay of the produced $J/\psi$.
That is, the angular distribution of decaying leptons in the process,
$J/\psi \rightarrow l^+l^-$, is given by
\begin{eqnarray} 
\frac{d\Gamma}{d\cos\theta_{l}^*} \propto \left[ 
1 + \alpha (z) ~\cos^2 \theta_l^* \right],
\end{eqnarray} 
with $\theta_l^*$ the angle between the lepton three-momentum in the
$J/\psi$ rest frame and the $J/\psi$ direction in the lab frame.
The relation between $\alpha$ and $\eta$ is given by the following equation
\begin{equation}
\alpha (z) = \frac{1-3\eta (z)}{1+\eta(z)}.
\end{equation}
Therefore, the unpolarized $J/\psi$ corresponds to $\eta=1/3$ ($\alpha =0$),
whereas the pure transverse polarization corresponds to 
$\eta=0$ ($\alpha=1$).

The dominant color-singlet contributions come from the following two
processes:
\begin{eqnarray}
  e^+ e^- &\rightarrow& (c\bar{c})[{^3S_1}]^{(1)} + g g      
\label{eq:eejgg}  \\
  e^+ e^- &\rightarrow& (c\bar{c})[{^3S_1}]^{(1)}  + c\bar{c}.  
\label{eq:eejcc}
\end{eqnarray}   
Among color-octet contributions, the following two processes are dominant.
\begin{eqnarray}
e^+ e^- &\rightarrow& (c\bar{c})[{^1S_0}~{\rm or} ~{^3P_J}]^{(8)}  + g 
\label{eq:eejg}\\
e^+ e^- &\rightarrow& (c\bar{c})[{^3S_1}]^{(8)}  + q\bar{q} 
\label{eq:eejqq} .
\end{eqnarray}
We show the angular momentum and spin quantum numbers of the $c\bar{c}$ 
in the spectroscopy notation, and the superscripts (1) and (8) represent 
its color structures.  These $c\bar{c} [ {^{2S+1}L_J} ]^{1,8}$ states
will eventually evolve into a physical $J/\psi$ by emitting/absorbing 
soft gluons, the probabilities for which are parametrized in terms of 
NRQCD matrix elements, $\langle  O_{1,8}^{J/\psi} ({^{2S+1}L_J})  \rangle$.  

We have calculated the cross section for the longitudinal $J/\psi$ as well 
as the total cross section for $J/\psi$ production in the $e^+ e^-$ 
annihilations, and the full expressions are given in the Appendix. 
The virtual $Z^0$ contributions were neglected, which should be a good 
approximation if $\sqrt{s}$ is far below $M_Z$ \footnote{See
Refs.~\cite{z0} \cite{z0_frag} for $Z^0 \rightarrow J/\psi + X$ at LEP,
Ref.~\cite{z0pol} for the $J/\psi$ polarization therein.}.  
Our results for the total production cross sections agree with the previous
results obtained in Refs.~\cite{cho} and \cite{fyuan} \footnote{There  are 
some typos in the overall factors of Eqs.(3.8),(A2a) and (A2b)
in Ref.~\cite{cho}.  
Their  results should be multiplied by a factor of 3.}.

For  numerical analyses, we use the following numbers for 
nonperturbative matrix elements that appear in the NRQCD factorization 
formula for the $J/\psi$ production cross sections :
\begin{eqnarray}
\langle {\mathcal O}^{J/\psi}_1(^3S_1) \rangle &=&0.73 ~\mbox{GeV}^3\\
\langle {\mathcal O}^{J/\psi}_8(^3S_1) \rangle &=&0.015 ~\mbox{GeV}^3\\
\langle {\mathcal O}^{J/\psi}_8(^1S_0) \rangle &\approx&10^{-2} 
~\mbox{GeV}^3\\
\langle {\mathcal O}^{J/\psi}_8(^3P_0) \rangle /m_c^2 &\approx&10^{-2} 
                 ~\mbox{GeV}^3.
\end{eqnarray}
We set $\alpha_s(2m_c) = 0.28$  with $m_c = 1.48$ GeV.

In Fig~1, we show the $J/\psi$ production cross sections for different 
$J/\psi$ production mechanisms as functions of the beam energy, 
$E_{\rm beam}$. 
At low electron-beam energies ($E_{\rm beam} < 10$ ~GeV) such as CLEO, 
the color-octet process ~(\ref{eq:eejg}) and the color-singlet process (4) 
dominate over other mechanisms.
As the beam energy increases, the cross section via (\ref{eq:eejgg})
decreases very rapidly, proportional to inverse 
fourth of the beam energy($\propto E_{\rm beam}^{-4}$).  
And if the beam energy is
greater than about $10$ GeV, the quark process, which decreases according to
the inverse square of the beam energy ($\propto E_{\rm beam}^{-2}$), 
dominates as shown in Fig~1. The hard gluon process Eq.~(\ref{eq:eejg}) 
dominates when the electron-beam energy is lower than about $10$~GeV, and 
the gluon fragmentation process Eq.~(\ref{eq:eejqq}) dominates when the 
electron-beam  energy is higher than $10$~GeV \cite{fyuan}. 

In Fig.~2, we show the energy distribution($d\sigma/dz$) of the prompt 
$J/\psi$ produced  at CLEO, where $E_{\rm beam} = 5.29$ GeV and $z$ is the 
energy fraction $E_{J/\psi}/E_{\rm beam}$  in $e^+e^-$ center of mass frame.  
As shown in Fig.~2,  the $J/\psi$'s produced  via the color-singlet gluon 
mode  (\ref{eq:eejgg}) are roughly three times more 
than those via the color-singlet charm-quark fragmentation (\ref{eq:eejcc}).  
The color-octet  contribution is suppressed relative to
the color-singlet processes except at the upper end point of
phase space, where the color-octet process (\ref{eq:eejg}) dominates 
\cite{Bra}.

In Fig.~3, we show the polarizations of $J/\psi$'s for each production 
mechanism. It turns out that each production mechanism leads to vastly 
different polarization of $J/\psi$'s. 
The color-singlet gluon mode (\ref{eq:eejgg})  makes longitudinally
polarized $J/\psi$'s, so that $\alpha(z)$ is negative in all the
range of $z$.  Especially, the high energy $J/\psi$'s made by 
(\ref{eq:eejgg})  are almost completely longitudinally polarized  
($\alpha \rightarrow -1$). 
The $J/\psi$'s produced via the color-octet mechanism (\ref{eq:eejg})
are  populated only at the  point region of phase space, 
since it is a two-body process at the parton level.  And they
are almost unpolarized ($\eta = 0.35$, or equivalently, 
$\alpha=-0.05$).  The $J/\psi$'s via the color-octet gluon fragmenation
process (\ref{eq:eejqq})  tend to be transversely polarized over all their 
energy range, and they become completely transversely polarized at the 
end-point region. However since this process is not dominant at CLEO energy,
we do not  consider it anymore. In Fig.~ 3, we can see that at the
upper end point of phase space, if the color-octet process (\ref{eq:eejg}) 
dominates, $\alpha \approx -0.05$, and if the color-singlet process 
(\ref{eq:eejgg}) dominates, $\alpha \approx -0.86$.  The $J/\psi$ 
polarization at the end point region at CLEO depends on the matrix elements 
of NRQCD as follows:
\begin{equation}
\alpha = \frac{-0.05 \langle {\mathcal O}^{J/\psi}_8(^1S_0) \rangle
-0.82 \langle {\mathcal O}^{J/\psi}_1(^3S_1) \rangle
-4.8 \langle {\mathcal O}^{J/\psi}_8(^3P_0) \rangle /m_c^2}
{19.0 \langle {\mathcal O}^{J/\psi}_8(^1S_0) \rangle
+0.94 \langle {\mathcal O}^{J/\psi}_1(^3S_1) \rangle
+72.0 \langle {\mathcal O}^{J/\psi}_8(^3P_0) \rangle/m_c^2 }\;.
\end{equation}
  So the precise measurement of the 
$J/\psi$ polarizations near the end point region at CLEO may provide
us with another information on NRQCD  matrix elements.     
In particular, any appreciable deviation from $\alpha ({\rm singlet}) = 
-0.86$ may be a signal of importance of color-octet mechanism. Finally
in Fig.~4, we show $\alpha(z)$ obtained by summing all the contributions from
various  $J/\psi$ production mechanisms at CLEO energy.

Finally, let us remark on the possible breakdown of NRQCD near the end point 
region of phase space that  was recently pointed out by Beneke, Rothstein 
and Wise \cite{Bene}.  The  channel that might have this problem is the 
$e^+ e^- \rightarrow (c\bar{c})[{^1S_0}~{\rm or} ~{^3P_J}]^{(8)}+g$ mode. If 
we consider the soft gluon emission during the evolution of the color-octet 
$(c\bar{c})^{(8)}$ states  into $J/\psi$, the $z$-distribution given in Ref.~
\cite{Bra} would be changed schematically to the following form :
\begin{eqnarray}
{d \sigma \over dz } (e^+e^-\rightarrow (c\bar{c})[^1S_0]^{(8)}+g)
&=&\int dy_E {d\Omega \over 4\pi} \delta(z-(1+\delta^2/4)-y_E)  
C_S {3 \over 4} (1+\cos^2\theta)
 F[^1 S_0^{(8)}](y_E), 
\end{eqnarray}
where $C_s$ is defined in the Appendix (see Eq.~(22)), 
and the shape function $F$ is given by \cite{Bene}
\begin{eqnarray}
F[^1 S_0^{(8)}](y_E) = \sum_{X}
\langle 0|\psi^\dagger T^a \chi|H+X\rangle
\langle H+X|\delta(y_E-(1-{\delta^2 \over 4}) i n\cdot\hat{D})
\chi^\dagger T^a \psi |0\rangle \;.
\end{eqnarray}
This shape function $F[^1 S_0^{(8)}]$ represents the distribution of 
energy fraction carried away by soft gluons during the hadronization of 
the color-octet $c\bar{c} ({^1S_0})^{(8)}$ pair into the physical $J/\psi$, 
in the $J/\psi$  rest frame \cite{Bene}.  However, $C_s$ is independent of 
the variable $z$, and  
\begin{equation}
\int dy_E F[^1 S_0^{(8)}](y_E) =\langle 0| {\mathcal O}^H_8(^1 S_0)|0 \rangle
\;.
\end{equation}
Therefore, the average over some small region near the end point gives 
just the NRQCD form.  This is different from the case of hadroproduction 
discussed in Ref.~\cite{Bene}.  The same argument applies to $c\bar{c}
[^3P_0]^{(8)}$ modes, and also to  the $J/\psi$ polarization as well.

In conclusion, we showed that the  $J/\psi$ polarization depends 
distinctively   on the $J/\psi$ production mechanisms.   At the upper
end point of phase space where $J/\psi$'s are  dominantly produced via
the  color-octet mechanism $e^+ e^- \rightarrow (c\bar{c})_{8} + g$, the 
measurement of the $J/\psi$ polarization can give another test of the  
color-octet mechanism and another constraint on the NRQCD matrix elements. 
We also argued that the possible breakdown of NRQCD near the phase space 
boundary does not affect our predictions made in this work.   
The ongoing data analysis at CLEO will shed light on the relevance of
the color-octet mechanism in $J/\psi$ productions through the measurements 
of the polar angle distribution and polarization of $J/\psi$'s in the 
$e^+ e^-$ annihilation at CLEO \cite{cleo}.   

\acknowledgements
We are grateful to  K.T. Chao, Victor Kim and  M.B. Wise for useful 
discussions.
This work is supported in part by KOSEF through CTP, BSRI-97-2418, and 
Nondirected Research by Ministry of Education (PK), and KOSEF Fellowship (JL).

\section{Appendix}
We list the analytic expressions for the differential cross sections of
total and longitudinal $J/\psi$ productions in terms of
\[
z = 2 E_{J/\psi} / \sqrt{s}, ~~{\rm and} ~~ \delta = 4 m_c / 
\sqrt{s}.
\]

\subsection{$e^+e^- \rightarrow c\bar{c}[^3S_1]^{(1)}+gg$}
\begin{eqnarray}
\label{eq:jgg}
\lefteqn{
\frac{d\sigma_{tot}}{d z}(e^+ e^- \rightarrow c\bar{c}[^3S_1]^{(1)} g g)
= \frac{8\pi}{81} \frac{(\alpha \alpha_s e_Q)^2}{\delta E_{\rm beam}^5}
\frac{<{\cal O}^\psi_1(^3S_1)>}{(2-z)^2  (2 z -\delta^2)^3}
}\nonumber\\
&\times& \Bigg\{
4 \Big[ - 16 z^3 + 2 z^2 (7 \delta^2 + 26)
- 6 z (\delta^2 + 2) (\delta^2 + 4)\nonumber \\
&&+ \delta^6 + 7 \delta^4 + 20 \delta^2 + 16 \Big]
(2 z -\delta^2) \sqrt{z^2-\delta^2}\nonumber\\
&+&\Big[2 z^2 (5 \delta^4 - 4 \delta^2 - 16)
+ 2 z \delta^2 ( - 3 \delta^4 - 4 \delta^2 + 40)\nonumber\\
&&- \delta ^2 (4-\delta^2) (\delta^4 + 8 \delta^2 + 4)\Big]
(4 z -4 -\delta^2)
\ln{\frac{2 z -\delta^2 +2 \sqrt{z^2 -\delta^2}}
{2 z -\delta^2 -2 \sqrt{z^2 -\delta^2}}}\Bigg\}.
\end{eqnarray}
\begin{eqnarray}
\label{eq:jgglong}
\lefteqn{
\frac{d\sigma_L}{d z}(e^+ e^- \rightarrow   c\bar{c}[^3S_1]^{(1)} g g)
= \frac{\pi}{324} \frac{(\alpha \alpha_s e_Q)^2}{\delta E_{\rm beam}^5}
\frac{<{\cal O}^\psi_1(^3S_1)>}{(2-z)^2  (2 z -\delta^2)^3 (z^2 -\delta^2)}
}\nonumber\\
&\times&\Bigg\{4\Big[
 128 z^4
+ 64 z^3 (  \delta^4 - 2 \delta^2 + 8)
- 32 z^2 (2 \delta^6 + 3 \delta^4 + 40 \delta^2 + 16) \nonumber\\
&&+ 8 z \delta^2 (  3 \delta^6 + 4 \delta^4 + 128 \delta ^2 + 128)
- \delta^4 (3 \delta^6 + 4 \delta^4 + 144 \delta^2 + 576)
\Big] (2 z -\delta^2) \sqrt{z^2 -\delta^2} \nonumber\\
&&-( 4 z - \delta^2- 4) \Big[
64 z^4 (\delta^4 + 4 \delta^2 + 16)
+ 32 z^3 \delta^2 ( - 3 \delta^4 - 12 \delta^2 - 80) \nonumber\\
&&+ 8 z^2 \delta^2 (9 \delta^6 + 8 \delta^4 + 256 \delta^2 - 64)
+ 8 z \delta^4 ( - 3 \delta^6 - 64 \delta^2 + 128) \nonumber\\
&& - \delta^6 (4-\delta^2)  (3 \delta^4 + 8 \delta^2 + 144)
\Big]
\ln{\frac{2 z -\delta^2 +2 \sqrt{z^2 -\delta^2}}
{2 z -\delta^2 -2 \sqrt{z^2 -\delta^2}}}\Bigg\}.
\end{eqnarray}

\subsection{$e^+e^- \rightarrow c\bar{c}[^3S_1]^{(1)}+c\bar{c}$}
\begin{eqnarray}
\lefteqn{
\frac{d\sigma_{tot}}{d z}(e^+ e^- \rightarrow c\bar{c}[^3S_1]^{(1)} c\bar{c})
= \frac{\pi}{486} \frac{(\alpha \alpha_s e_Q)^2}{\delta^3 E_{\rm beam}^5}
\frac{<{\cal O}^\psi_1(^3S_1)>}{z^3 (2 -z)^6}
}\nonumber\\
&\times&\Bigg\{4 z \sqrt{\frac{(1-z) (z^2 -\delta^2)}{4 +\delta^2 -4 z}}
\Big[
1280 z^8
+ 32 z^7 (\delta^2 - 336) \nonumber\\
&&+ 8 z^6 ( - 15 \delta^4 + 20 \delta^2 + 4512) \nonumber\\
&&+ 4 z^5 (3 \delta^6 + 136 \delta^4 - 512 \delta^2 - 13312)\nonumber\\
&&+ z^4 ( - 3 \delta^8 + 12 \delta^6 - 672 \delta^4 + 1792 \delta^2 + 38912) \nonumber\\
&&+ 4 z^3 (\delta^8 - 12 \delta^6 - 288 \delta^4 - 2432 \delta^2 - 4096) \nonumber\\
&&-16 z^2 ( \delta^2 + 4) (\delta^6 - 42 \delta^4 - 328 \delta^2 - 64) \nonumber\\
&&+ 16 z \delta^2 ( - 3 \delta^6 - 88 \delta^4 - 576 \delta^2 - 1024) \nonumber\\
&&+ 16 \delta^2 (\delta^2 + 4) (\delta^2 + 8) (3 \delta^2 + 8 )
\Big]\nonumber\\
&&+\Big[
- 512 z^5 + 8 z^4 (7 \delta^2 + 104)
+ 8 z^3 \delta^2 ( - 5 \delta^2 - 4)\nonumber\\
&&+ z^2 (3 \delta^6 + 32 \delta^4 - 32 \delta^2 - 256)
+ 128 z \delta^2 ( - \delta^2 - 4)
+ 4 \delta^2 (3 \delta^4 + 32 \delta^2 + 64)\nonumber\\
&&\Big]
\delta^2 (2-z)^4
\ln\frac{z\sqrt{4+\delta^2-4 z} +2\sqrt{(1-z)(z^2-\delta^2)}}
{z\sqrt{4+\delta^2-4 z} -2\sqrt{(1-z)(z^2-\delta^2)}}
\Bigg\}.
\end{eqnarray}
\begin{eqnarray}
\lefteqn{
\frac{d\sigma_L}{d z}(e^+ e^- \rightarrow c\bar{c}[^3S_1]^{(1)} c\bar{c})
= \frac{\pi}{1458} \frac{(\alpha \alpha_s e_Q)^2}{\delta^3 E_{\rm beam}^5}
\frac{<{\cal O}^\psi_1(^3S_1)>}{z^3 (2 -z)^6 (z^2 -\delta^2)}
}\nonumber\\
&\times& \Bigg\{ 4z\sqrt{\frac{(1-z)(z^2-\delta^2)}{4+\delta^2-4z}}
\Big[ 768 z^{10}
+ 96 z^9 (\delta^2 - 80) \nonumber\\
&&+ 8 z^8 (9 \delta^4 - 116 \delta^2 + 3680) \nonumber\\
&&+ 4 z^7 ( - 39 \delta^6 - 204 \delta^4 + 1696 \delta^2 - 11776) \nonumber\\
&&+ 2 z^6 (15 \delta^8 + 732 \delta^6 + 1088 \delta^4 - 15104 \delta^2 + 18432) \nonumber\\
&&+ 16 z^5 ( - 17 \delta^8 - 304 \delta^6 - 112 \delta^4 + 2240 \delta^2 - 1024)
 \nonumber\\
&&+ z^4 (3 \delta^10 + 800 \delta^8 + 8272 \delta^6 + 9600 \delta^4 + 512 \delta^2 + 4096) \nonumber\\
&&+ 4 z^3 \delta^2 ( - \delta^8 - 296 \delta^6 - 1440 \delta^4 - 1984 \delta^2 -
 5120) \nonumber\\
&&+ 16 z^2 \delta^2 (\delta^8 - 4 \delta^6 - 456 \delta^4 - 800 \delta^2 + 512) \nonumber\\
&&+ 48 z \delta^4 (\delta^6 + 40 \delta^4 + 272 \delta^2 + 384) \nonumber\\
&&+ 48 \delta^4 ( - \delta^6 - 24 \delta^4 - 112 \delta^2 - 128) \Big]\nonumber\\
&&-3\delta^2(2-z)^4
\ln\frac{z\sqrt{4+\delta^2-4 z} +2\sqrt{(1-z)(z^2-\delta^2)}}
{z\sqrt{4+\delta^2-4 z} -2\sqrt{(1-z)(z^2-\delta^2)}} \Big[\nonumber\\
&&   8 z^6 (\delta^2 - 40)
+ 28 z^5 \delta^2 ( - \delta^2 + 8)
+ 2 z^4 (5 \delta^6 + 24 \delta^4 + 144 \delta^2 + 128) \nonumber\\
&&+ 4 z^3 \delta^4 ( - 7 \delta^2 - 36)
+ z^2 \delta^2 (\delta^6 + 12 \delta^4 - 96 \delta^2 - 256) \nonumber\\
&&+ 16 z \delta^4 ( - 3 \delta^2 - 8)
+ 4 \delta^4 (\delta^4 + 20 \delta^2 + 32) \Big] \Bigg\}.
\end{eqnarray}

\subsection{$e^+e^-\rightarrow c\bar{c}[^3S_1]^{(8)}+q\bar{q}$}
\begin{eqnarray}
\lefteqn{
\frac{d\sigma_{tot}}{dz}(e^+e^- \rightarrow c\bar{c}[^3S_1]^{(8)} q \bar{q}) =
{\pi \over 72} \frac{\langle {\mathcal O}_8^\psi(^3S_1)\rangle}{m_c^3 
E_{\rm beam}^2}
(\alpha \alpha_s e_Q)^2
}\nonumber\\
&\times&\Bigg[\log\left(
       \frac{z+\sqrt{z^2-\delta^2}}{z-\sqrt{z^2-\delta^2}}\right)
      \left\{4z -2(4+\delta^2) +\frac{(4+\delta^2)^2}{2z}\right\}
    -8 \sqrt{z^2-\delta^2} \Bigg].
\end{eqnarray}
\begin{eqnarray}
\lefteqn{
\frac{d\sigma_L}{dz}(e^+e^- \rightarrow c\bar{c}[^3S_1]^{(8)} q \bar{q}) =
{\pi \over 72} \frac{\langle {\mathcal O}_8^\psi(^3S_1)\rangle}{m_c^3 
E_{\rm beam}^2}
(\alpha \alpha_s e_Q)^2
}\nonumber\\
&\times&\frac{(4z-(4+\delta^2))(4+\delta^2)}{2(z^2-\delta^2)}
 \Bigg[\log\left(
       \frac{z+\sqrt{z^2-\delta^2}}{z-\sqrt{z^2-\delta^2}}\right)
         /2z -\sqrt{z^2-\delta^2}\Bigg].
\end{eqnarray}

\subsection{$e^+e^-\rightarrow c\bar{c}[^1S_0 ~{\rm or}~ ^3P_0]^{(8)}+g$}
\begin{eqnarray}
\sigma(e^+e^- \rightarrow c\bar{c}[^1S_0 ~{\rm or}~ ^3P_0]^{(8)} g) &=&
 C_S <{\mathcal O}_8^\psi(^1 S_0)>
+C_P <{\mathcal O}_8^\psi(^3 P_0)>, ~{\rm where} 
\\
C_S &=& \frac{8 \pi^2 \alpha^2 e_Q^2 \alpha_s}{3 s^2 m_c}
       (4 -\delta^2),
\nonumber  \\
C_S^{long} &=& C_S/3,
\\
C_P &=& \frac{8 \pi^2 \alpha^2 e_Q^2 \alpha_s}{3 s^2 m_c^3}
      \frac{48 + 8 \delta^2 +7 \delta^4}{4 -\delta^2},
\nonumber\\
C_P^{long} &=& \frac{8 \pi^2 \alpha^2 e_Q^2 \alpha_s}{3 s^2 m_c^3}
      \frac{(4+\delta^2)^2}{4 -\delta^2}.
\nonumber 
\end{eqnarray}
Here, $C_{S,P}^{long}$ correspond to the longitudinal $J/\psi$ production.



\newpage
\begin{center}
{\Large\bf FIGURE CAPTIONS}
\end{center}
\noindent Fig.~1
The  cross section for each mode  versus electron beam energy:
$e^+e^- \rightarrow J/\psi+q\bar{q}$ (solid),
$e^+e^- \rightarrow J/\psi+c\bar{c}$ (long-dashed),
$e^+e^- \rightarrow J/\psi+gg$ (short-dashed),  and
$e^+e^- \rightarrow J/\psi+g$ (dotted).
\vskip1cm
\noindent Fig.~2
Energy spectra  of $J/\psi$'s at CLEO energy for each mode : 
$e^+e^-\rightarrow J/\psi +gg$,
$e^+e^-\rightarrow J/\psi +c\bar{c}$,
$e^+e^-\rightarrow J/\psi +q\bar{q}$, and
$e^+e^-\rightarrow J/\psi +g$.
\vskip1cm
\noindent Fig.~3
$\alpha(z)$ for each mechanism at CLEO energy.
\vskip1cm
\noindent Fig.~4
$\alpha(z) $ for the sum of all modes at CLEO energy. 
\vskip1cm

\newpage
\begin{figure}[!t]
\begin{center}
\includegraphics[height=9.2cm,width=13cm]{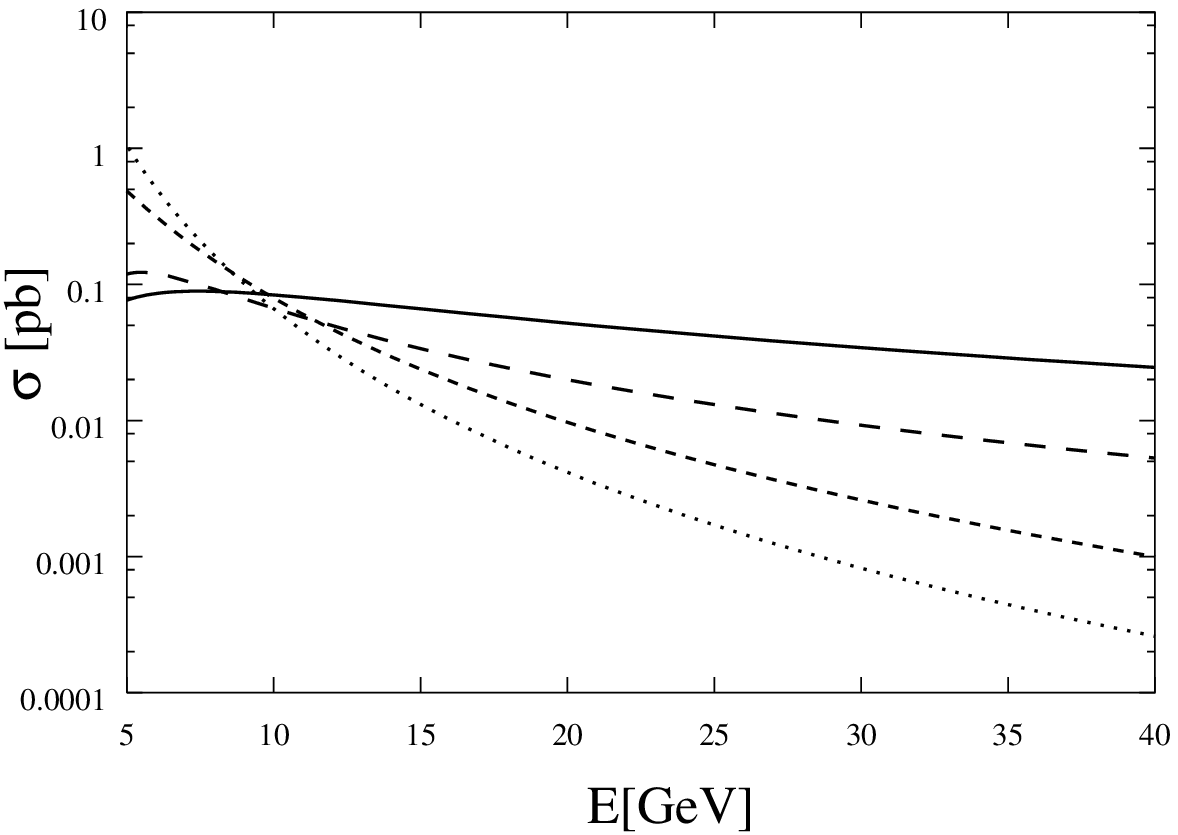}
\caption{
}
\end{center}
\end{figure}
\begin{figure}[!b]
\begin{center}
\includegraphics[height=8cm,width=11cm]{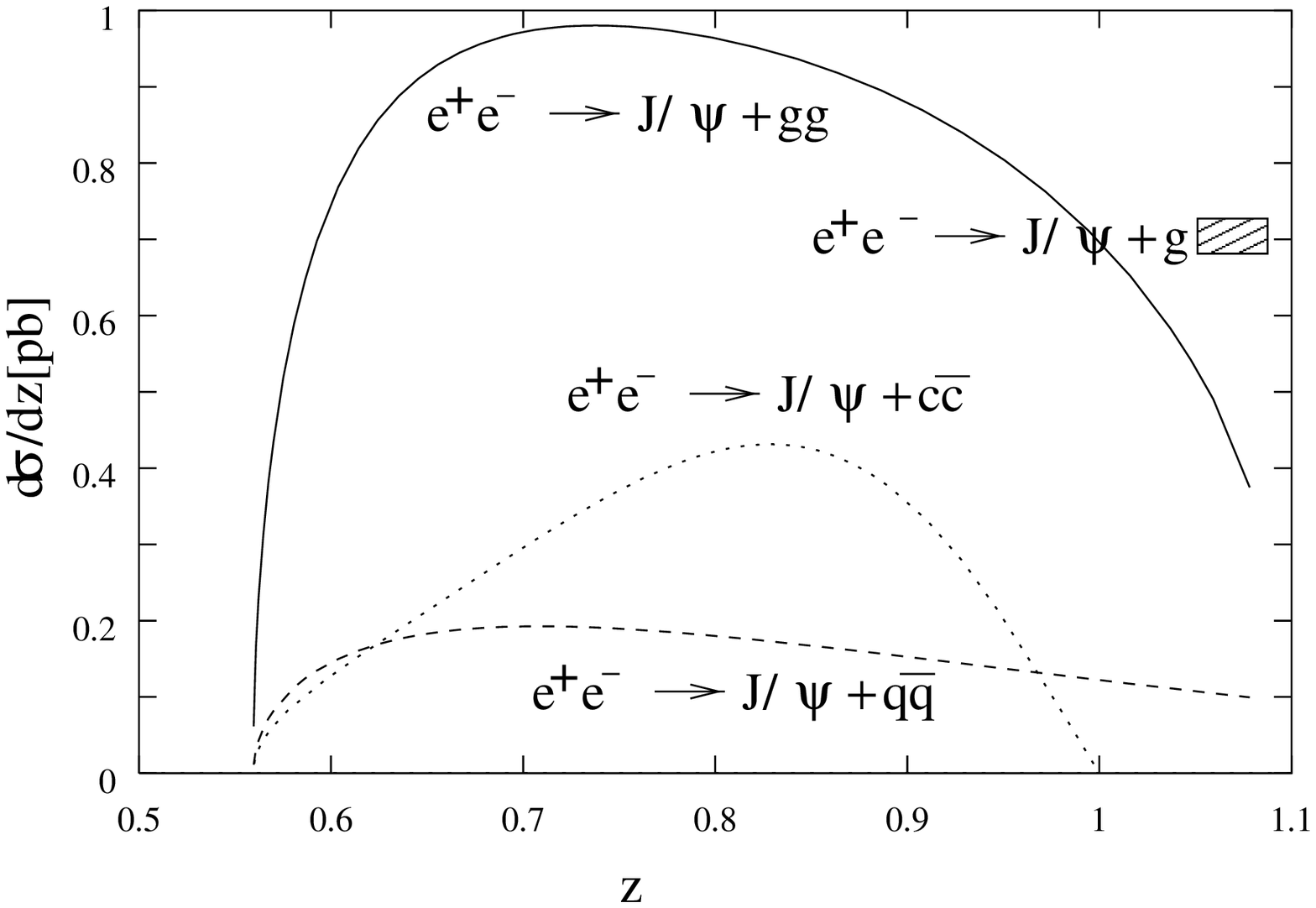}
\caption{
}
\end{center}
\end{figure}
\begin{figure}[!t]
\begin{center}
\includegraphics[height=8cm,width=11cm]{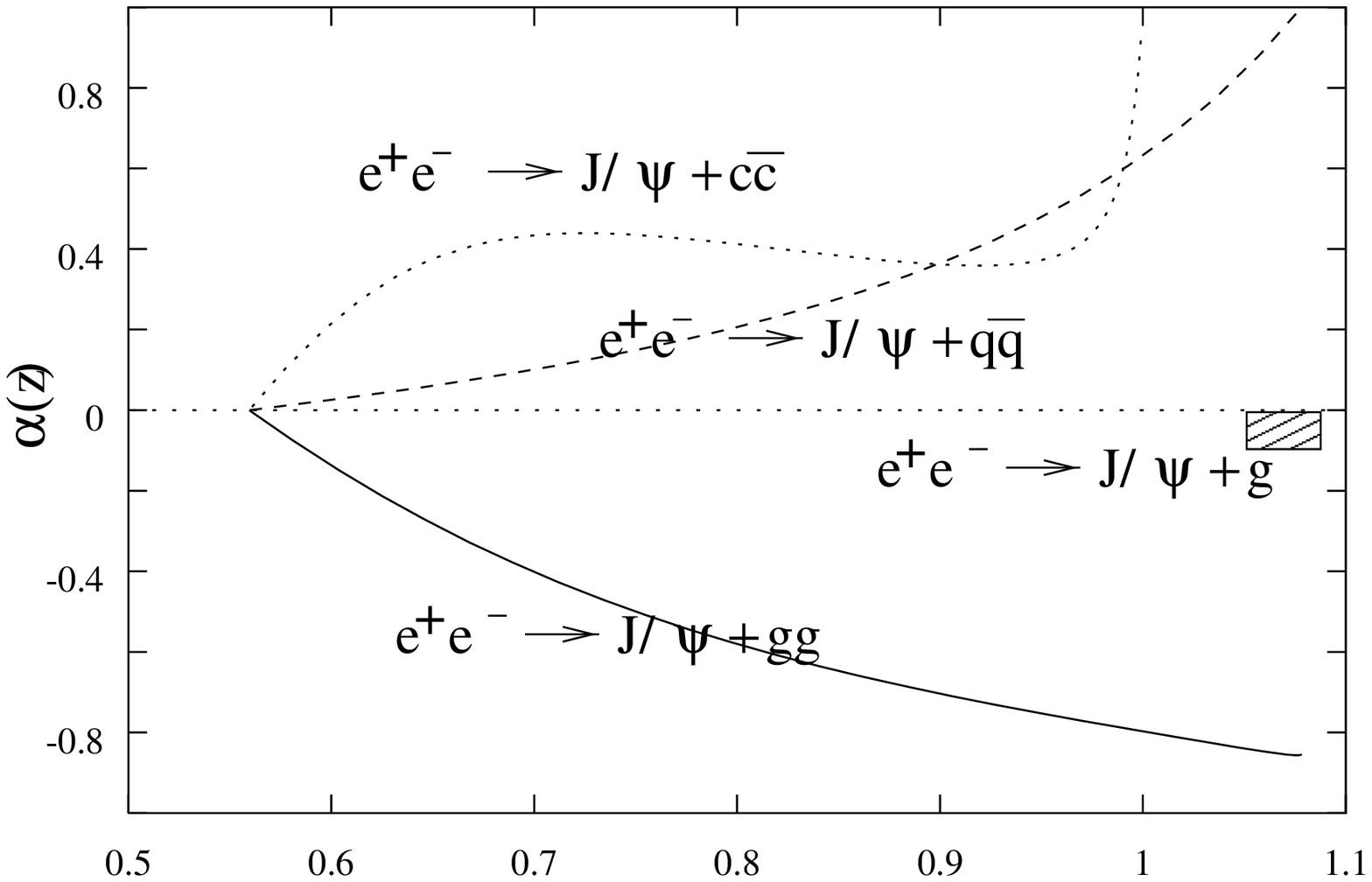}
\caption{
}
\end{center}
\end{figure}
\begin{figure}[!h]
\begin{center}
\includegraphics[height=8cm,width=11cm]{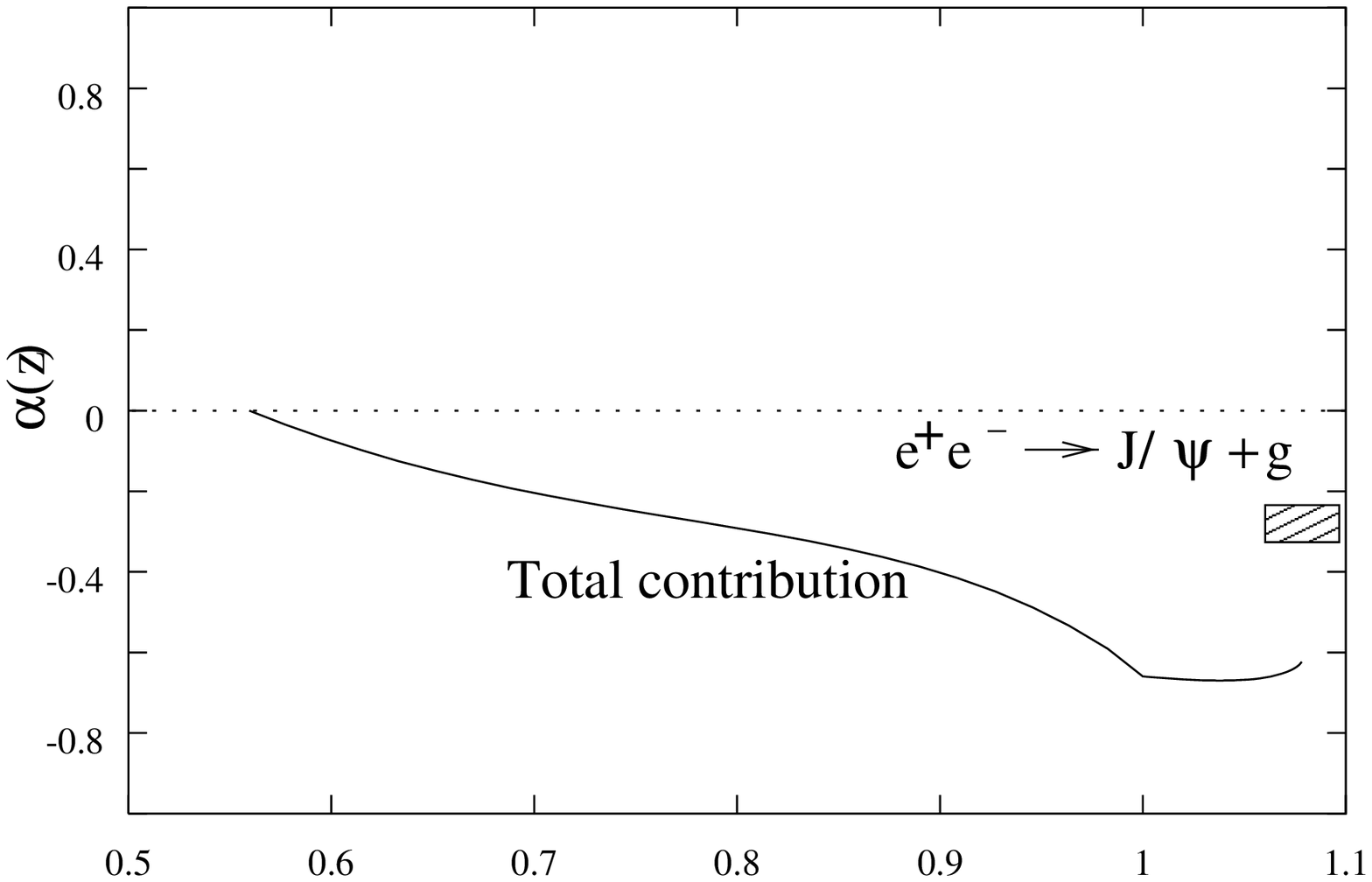}
\caption{
}
\end{center}
\end{figure}
\end{document}